# EOCAS: Energy-Oriented Computing Architecture Simulator for SNN Training

Yunhao Ma, Wanyi Jia, Yanyu Lin, Wenjie Lin, Xueke Zhu, Huihui Zhou and Fengwei An, *Member, IEEE*

*Abstract*—With the growing demand for intelligent computing, neuromorphic computing, a paradigm that mimics the structure and functionality of the human brain, offers a promising approach to developing new high-efficiency intelligent computing systems. Spiking Neural Networks (SNNs), the foundation of neuromorphic computing, have garnered significant attention due to their unique potential in energy efficiency and biomimetic neural processing. However, current hardware development for efficient SNN training lags significantly. No systematic energy evaluation methods exist for SNN training tasks. Therefore, this paper proposes an Energy-Oriented Computing Architecture Simulator (EOCAS) for SNN training to identify the optimal architecture. forward propagation, backward propagation, and weight update based on gradient. EOCAS combines design representation, energy efficiency evaluation and dataflow consideration during each simulation to identify the optimal architecture. Under the guidance of EOCAS, we implement the optimal hardware architecture through Verilog HDL and achieve low energy consumption using Synopsys Design Compiler with TSMC-28nm technology library under typical parameters. Compared with several State-Of-The-Art (SOTA) DNN and SNN works, our hardware architecture outstands others in various criteria.

## I. INTRODUCTION

WITH the growing demand for intelligent computing, traditional architectures are increasingly challenged in addressing energy efficiency concerns [1]. Neuromorphic computing, a paradigm that mimics the structure and functionality of the human brain, offers a promising approach to developing new high-efficiency intelligent computing systems [2]. As the foundation of neuromorphic computing, Spiking Neural Networks (SNNs) have garnered significant attention due to their unique potential in energy efficiency and biomimetic neural processing [3]. With deep SNN models achieving classification accuracy comparable to traditional Deep Neural Networks (DNNs), they demonstrate substantial potential for enabling efficient intelligent computing.

However, current hardware development for efficient SNN training lags significantly. Existing near-memory computing chips such as TrueNorth [4], Loihi [5], and Tianjic [6] primarily support feedforward computation and local learning methods like spike-timing-dependent plasticity (STDP). [19] presents a hardware-efficient and scalable strategy for implementing large-scale, biologically realistic spiking neural networks. [20] introduces a biologically inspired cognitive supercomputing system designed to emulate large-scale brain-like cognition. However, these works lack the capability for deep SNN training based on global optimization approaches like error Back Propagation. Accelerators like [7] for SNN are also supportive only for inferencing, lacking the flexibility for local training tasks. These phenomenon forces current SNN training to rely heavily on Graphic Processing Units (GPUs). Due to the GPUs' lack of specialized optimizations for SNN-specific features such as multi-step computation and spike convolution, SNN training consumes more memory, requires longer training times, and exhibits lower training/inference energy efficiency compared to DNN training on GPUs.

The systematic energy consumption evaluation simulation is a foundation for supporting the rapid iteration of high-energy-efficiency computing architecture development. Several relatively mature methods and simulators have already been established to make efficient DNN hardware design such as TimesLoop [8], ZigZag [9] and LLMCompass [18]. They proposal have very low error rate compared to real hardware (e.g. 10.9% for LLMCompass). However, these simulators lack dedicated optimization for deep SNNs, such as binary operations like spike-based convolution. Recent studies have proposed SNN training architectures, including SATA [15] and H2Learn [10]. SATA builds upon the 8-bit ANN inference accelerator Eyeriss [21] by designing a sparse SNN training architecture along with an energy evaluation mechanism, mainly targeting shallow models such as VGG-5. H2Learn relies on external computing platforms to precompute the contents of the look-up tables. However, neither SATA nor H2Learn provides hardware implementations.

This paper proposes an Energy-Oriented Computing Architecture Simulator (EOCAS) for SNN training to identify the optimal architecture. EOCAS considers unique hardware design representations and computation patterns to support energy optimization in various architectures. The main contributions of this paper are as follows:

This work is supported by the key project of the Pengcheng Laboratory (PCL2024AS204) and the Shenzhen Science and Technology Program (Grant No. KJZD20230923113300002 and JCYJ20241206180301001). Corresponding authors: Huihui Zhou.

Yunhao Ma is with the Southern University of Science and Technology, and Pengcheng Laboratory, Shenzhen, China.

Wanyi Jia is with the Shenzhen Institutes of Advanced Technology, the Chinese Academy of Sciences, University of Chinese Academy of Sciences, and Pengcheng Laboratory, Shenzhen, China.

Yanyu Lin, Wenjie Lin, Xueke Zhu are with the Pengcheng Laboratory, Shenzhen, Shenzhen, China.

Huihui Zhou is with the Shenzhen Institute of Advanced Technology, and Pengcheng Laboratory, Shenzhen, China. (e-mail: zhouhh@pcl.ac.cn)

Fengwei An is with the Southern University of Science and Technology, and Pengcheng Laboratory, Shenzhen, China. (e-mail: anfw@sustech.edu.cn)

Yunhao Ma and Wanyi Jia contribute the same. (Equal Contribution)

1. We propose the Energy-Oriented Computing Architecture Simulator (EOCAS) for deep SNN training tasks considering forward propagation, backward propagation, and weight update based on gradient. EOCAS combines design representation and energy efficiency evaluation during each simulation to identify the optimal architecture.

2. We present a special dataflow named advanced WS in the convolution process for deep SNN training. This dataflow is more aligned with the characteristics of convolutions, as it maps the convolution process to a structure where multiple presynaptic neuron groups connect to a single postsynaptic group, with its performance evaluation outstanding commonly used dataflows.

3. Guided by EOCAS, we implement the entire hardware architecture containing the forward (FWD) to handle forward propagation and the backward (BWD) to resolve backward propagation, and weight updating for SNN training. Compared with several State-Of-The-Art (SOTA) DNN and SNN works, our hardware architecture outstands others in various criteria.

## II. DEEP SNN TRAINING AGORITHM

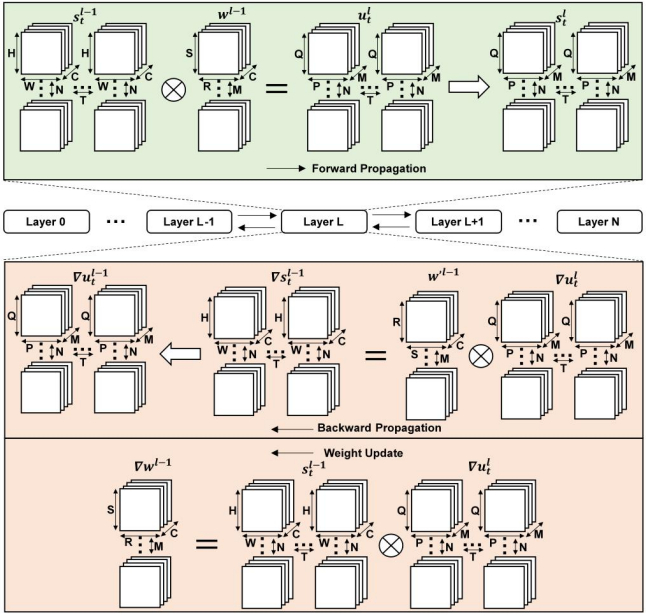

Fig. 1. The overall SNN training task.

### A. Deep SNN Training Tasks

For an L-layer deep SNN model $\mathcal{N} = \{L_l = \{u^l, w^l, s^l, ConvFP^l, \nabla u^l, w'^l, \nabla s^l, ConvBP^l, \nabla w^l\}; l = 1, 2, ..., L\}$, every layer $L_l$ is consist of neuron model LIF with $T$ timestep. L-layer neurons also contain the membrane potential $u^l = \{u_t^l \in \mathbb{R}^{B \times C^l \times H^l \times W^l}; t = 1, 2, ..., T\}$, the spike signal $s^l = \{s_t^l \in \varphi^{B \times C^l \times H^l \times W^l}; t = 1, 2, ..., T\}$, convolution $ConvFP^l = \{ConvFP_t^l \in \mathbb{R}^{B \times C^l \times H^l \times W^l}; t = 1, 2, ..., T\}$ and weight $w^l$ with its gradient $\nabla w^l \in \mathbb{R}^{M^l \times C^l \times R^l \times S^l}$ respectively. Additionally, the membrane potential gradient $\nabla u^l$, spike gradient $\nabla s^l$, transpose of weights, and membrane potential gradient convolution of L-layer neurons are the followings: $\nabla u^l = \{\nabla u_t^l \in \mathbb{R}^{B \times C^l \times H^l \times W^l}; t = 1, 2, ..., T\}$, $\nabla s^l = \{\nabla s_t^l \in \mathbb{R}^{B \times C^l \times H^l \times W^l}; t = 1, 2, ..., T\}$, $w'^l \in \mathbb{R}^{C^l \times M^l \times S^l \times R^l}$, and $ConvBP^l = \{ConvBP_t^l \in \mathbb{R}^{B \times C^l \times H^l \times W^l}; t = 1, 2, ..., T\}$. Among these parameters, $\varphi \in \{0,1\}$, $\mathbb{R}$ is a set of real numbers, and $C^l$, $H^l$, and $W^l$ are the channel, length, and width of the feature map of L-layer, respectively, $M^l$, $R^l$, and $S^l$ are the number, length, and width of the convolution kernel $w^l$, and $B$ is the batch size of a training.

### B. SNN Training Algorithm Based on FP, BP and WG

The training process is shown in Fig.1. The FP process of the SNN training is in (1) – (3):

$$u_t^l = \alpha u_{t-1}^l \odot (1 - s_{t-1}^l) + ConvFP_t^l \quad (1)$$
$$ConvFP_t^l = s_t^{l-1} \otimes w^{l-1} \quad (2)$$
$$s_t^l = f(u_t^l) = \begin{cases} 1 & u_t^l \geq th_f \\ 0 & else \end{cases} \quad (3)$$

Where $\odot$ denotes element-wise multiplication of two matrices of the same shape, $\otimes$ denotes the convolution operation, and $t$ represents the time step. The potential update includes both temporal and spatial components: the temporal component is determined by the previous time step's potential, spike, and leakage factor $\alpha$, while the spatial component is determined by the weighted accumulation of spikes of the prior layer's neurons. When the potential exceeds the threshold $th_f$, the neuron fires and resets the membrane potential. Here, $f(u_t^l)$ is a step function, where if $u_t^l \geq th_f$, then $f(u_t^l) = 1$; otherwise, $f(u_t^l) = 0$. Other operations in the FP are element-wised, with several orders of magnitude less computational than the spike convolution operation, which can be ignored. If the sparsity of the L-layer spike convolution is $Spar^l \in [0,1]$, then the spike Multiplexer (Mux) and floating point 16 (FP16) Add operands required for the layer L-1 spike convolution are the following, where $M^{l-1}$ equals to $C^l$.

$$Mux_{ConvFP}^l = B \times T \times C^{l-1} \times H^l \times W^l \times M^{l-1} \times R^{l-1} \times S^{l-1} \quad (4)$$
$$Add_{ConvFP}^l = B \times T \times C^{l-1} \times H^l \times W^l \times M^{l-1} \times R^{l-1} \times S^{l-1} \times Spar^l \quad (5)$$

Potential gradient in BP in SNN model $\nabla u_t^l \in \mathbb{R}^{B \times C^l \times H^l \times W^l}$ is defined as:

$$\nabla u_t^l = \alpha \nabla u_{t+1}^l \odot (1 - s_t^l) + \beta \nabla s_t^l \odot f'(u_t^l) \quad (6)$$

The derivative of the step function $f'(u_t^l)$ is approximated by the spike curve. If $th_l \leq u_t^l \leq th_r$, $f'(u_t^l) = 1$, otherwise $f'(u_t^l) = 0$. The spike gradient $\nabla s_t^l$ is defined as:

$$\nabla s_t^l = -\alpha \nabla u_{t+1}^l \odot u_t^l + ConvBP_t^l \quad (7)$$
$$ConvBP_t^l = \nabla u_t^{l+1} \otimes w'^l \quad (8)$$

The spike gradient $\nabla s_t^l$ is also composed of temporal and spatial components. The temporal component is determined by the potential gradient at the next time step and the current time step's potential. In contrast, the spatial component is determined by the weighted accumulation of potential gradients from the next layer. Since both the membrane potential gradient and the weight are FP16, the FP16 Multiplication (Mul) and FP16 Add operands required for the convolution of the L-layer membrane potential gradient are the following.

$$Mul^l_{ConvBP} = Add^l_{ConvBP}$$
$$= B \times T \times C^{l+1} \times H^{l+1} \times W^{l+1} \times C^l \times R^l \times S^l \quad (9)$$

For WG, $\nabla w^l$ is calculated based on the membrane potential gradient of the current layer and the spike output signal of layer $L$-1 obtained during the forward computation.

$$\nabla w^l = \sum_t \nabla u^l_t \otimes s^{l-1}_t \quad (10)$$

The basic computational operations for the weight gradient are also spike Mux and FP16 Add. The sparsity of the expanded convolution of the L-layer spike is $Spar^l$, and according to the above correspondence, the Mux and Add operands required for the weight gradient of L-layer are the following, where $M^l$ equals to $C^{l+1}$:

$$Mux^l_{WG} = B \times T \times R^l \times S^l \times M^l \times C^l \times H^{l+1} \times W^{l+1} \quad (11)$$
$$Add^l_{WG} = B \times T \times R^l \times S^l \times M^l$$
$$\times (C^l \times H^{l+1} \times Spar^l \times W^{l+1} + 1) \quad (12)$$

## III. EOCAS OVERALL FRAMEWORK

### A. Simulation Prerequisite

For the main convolution calculations in the above training tasks (formula (2), (8), and (10)), we propose a general SNN training near-memory computing architecture, including FP and BP core, shown in Fig. 2. The In both cores, the registers for spike or FP16 are needed, settled as a default settlement. Configurable components are divided into computation arrays and memory blocks. Besides, the computation and memory access of "the soma unit" and "the grad unit" are static, due to their immutable behavior during training, which will be discussed in subsection D.

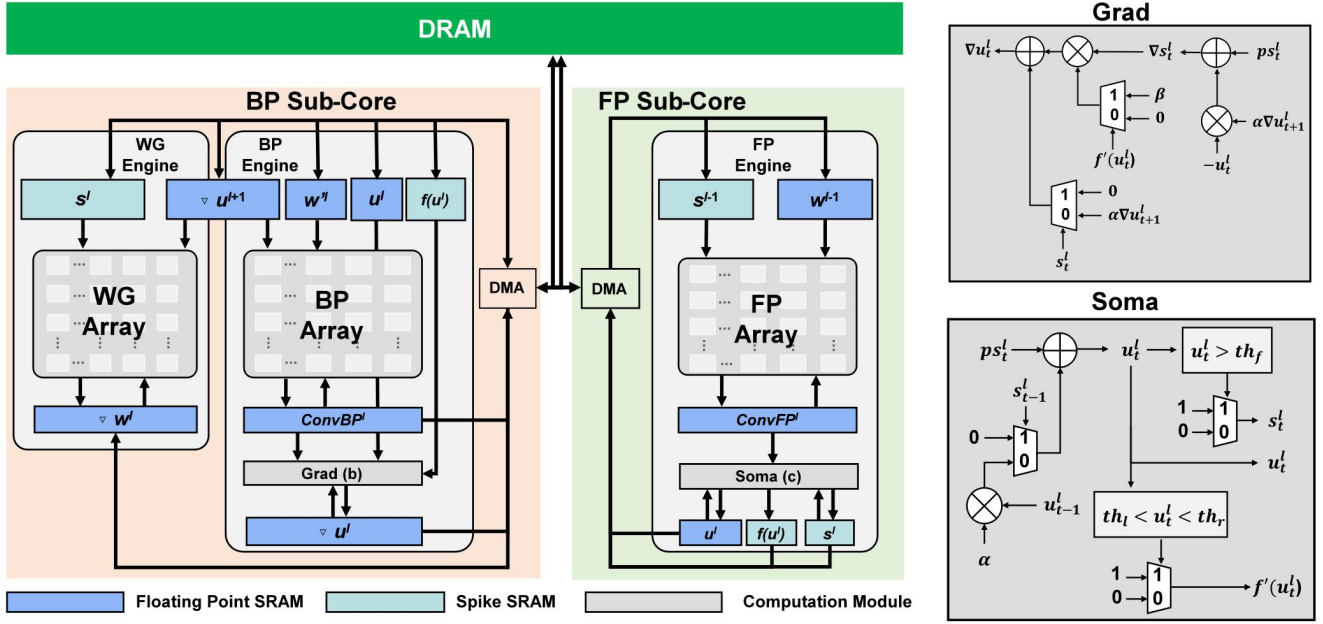

**Fig. 2.** A general SNN training computing architecture.

TABLE I
ENERGY CONSUMPTION AND REUSE FACTOR FOR EACH VARIABLE IN CONVOLUTIONS

| Type | Registers | | | | SRAM | | | | | | | | DRAM | | | |
|---|---|---|---|---|---|---|---|---|---|---|---|---|---|---|---|---|
| | W2S | | R2S | | R2R | | W2R | | W2D | | R2D | | R2S | | W2S | |
| | Reuse Factor (RF) | Energy | RF | Energy | RF | Energy | RF | Energy | RF | Energy | RF | Energy | RF | Energy | RF | Energy |
| $s^{l-1}_t$ | $RU_1$ | $r^w_0$ | —— | | $RU_1$ | $s^r_0$ | —— | | $RU_2$ | $s^w_0$ | —— | | $RU_2$ | $m^r_0$ | —— | |
| $w^{l-1}$ | $RU_3$ | $r^w_1$ | —— | | $RU_3$ | $s^r_1$ | —— | | $RU_4$ | $s^w_1$ | —— | | $RU_4$ | $m^r_0$ | —— | |
| $ConvFP^l$ | —— | | $RU_5$ | $r^r_1$ | —— | | $RU_5$ | $s^w_2$ | —— | | $RU_6$ | $s^r_2$ | —— | | $RU_6$ | $m^w_0$ |
| $\nabla u^{l+1}_t$ | $RU_7$ | $r^w_1$ | —— | | $RU_7$ | $s^r_3$ | —— | | $RU_8$ | $s^w_3$ | —— | | $RU_8$ | $m^r_0$ | —— | |
| $w'^l$ | $RU_9$ | $r^w_1$ | —— | | $RU_9$ | $s^r_4$ | —— | | $RU_{10}$ | $s^w_4$ | —— | | $RU_{10}$ | $m^r_0$ | —— | |
| $ConvBP^l$ | —— | | $RU_{11}$ | $r^r_1$ | —— | | $RU_{11}$ | $s^w_5$ | —— | | $RU_{12}$ | $s^r_5$ | —— | | $RU_{12}$ | $m^w_0$ |
| $\nabla u^{l+1}_t$ | $RU_{13}$ | $r^w_1$ | —— | | $RU_{13}$ | $s^r_3$ | —— | | $RU_{14}$ | $s^w_3$ | —— | | $RU_{14}$ | $m^r_0$ | —— | |
| $s^l_t$ | $RU_{15}$ | $r^w_0$ | —— | | $RU_{15}$ | $s^r_6$ | —— | | $RU_{16}$ | $s^w_6$ | —— | | $RU_{16}$ | $m^r_0$ | —— | |
| $\nabla w^l$ | —— | | $RU_{17}$ | $r^r_1$ | —— | | $RU_{17}$ | $s^w_7$ | —— | | $RU_{18}$ | $s^r_7$ | —— | | $RU_{18}$ | $m^w_0$ |

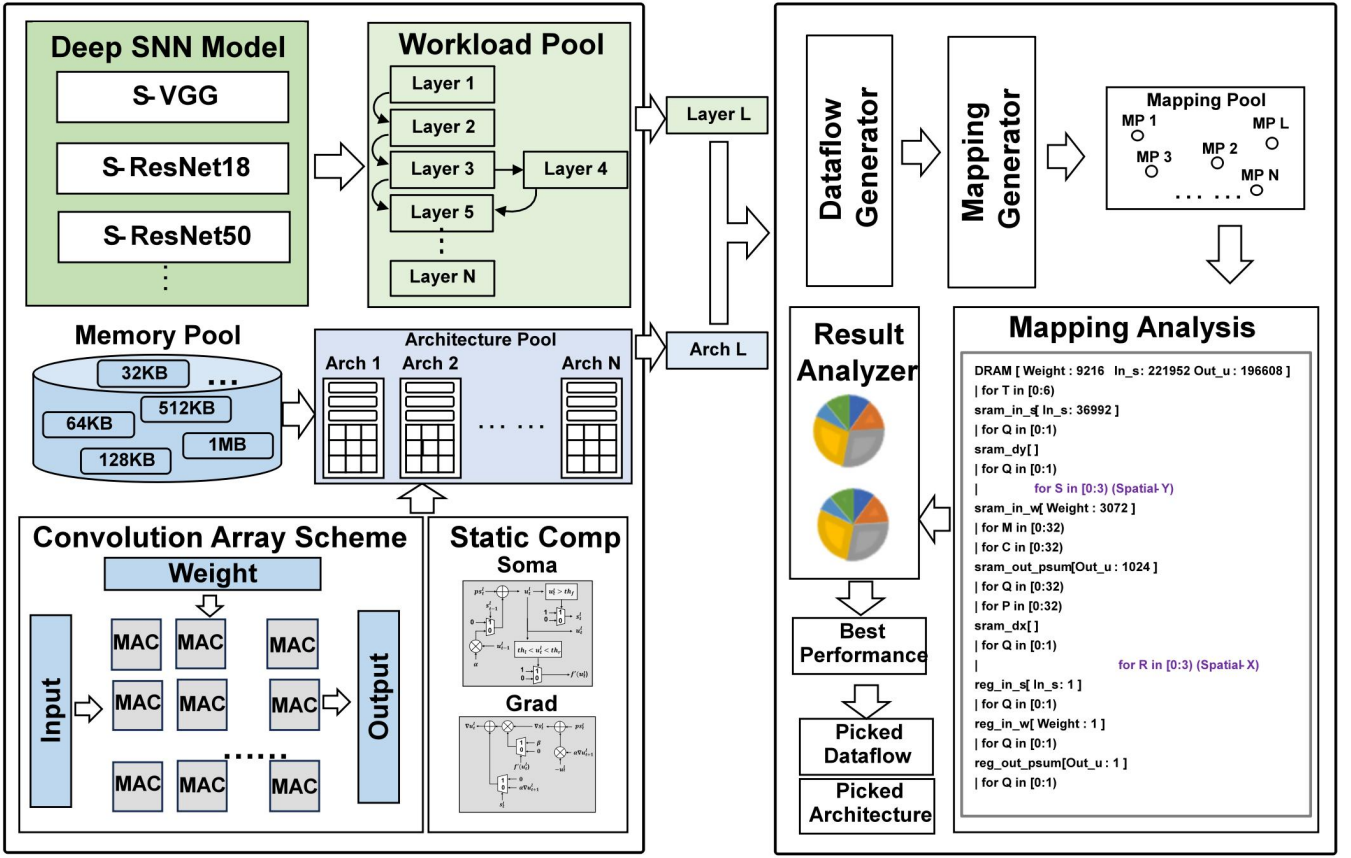

**Fig. 3.** The overall framework of EOCAS.

(1) For **FP** computation core, computation array consists of $E \times F$ Mux-Add units, where each Mux-Add unit consists of Mux, FP16 accumulator, 1-bit register for spike, and 16-bit registers for partial sum and weight. If the Mux is 0, the current accumulation is skipped. Each column of the Mux-Add array is configured with a column FP16 adder accumulator, which accumulates the partial sum outputs from the Mux-Add units in each column. Finally, a row FP16 adder accumulator aggregates the partial sum from all columns to obtain the convolution value or partial sum, completing all convolutions to get $ConvFP^l$. For FP memory, the on-chip storage mainly consists of three Static Random-Access Memory (SRAM) blocks storing $s^{l-1}$, $w^{l-1}$, $ConvFP^l$ respectively. These are called the spike SRAM ($V_1$), weight SRAM ($V_2$), and convolution SRAM ($V_3$).

(2) For **BP** computation core which contains BP computation and WG computation, the array consists of $E \times F$ Multiply (Mul)-Add units for FP16 MAC. Similar to FP, they also use a row adder to obtain the final sums, named $ConvBP^l$ and $WG^l$ respectively. For BP memory, the on-chip storage mainly consists of three SRAM blocks storing $\nabla u^{l+1}$, $w'^l$, $ConvBP^l$ respectively. These are called the delta-potential SRAM ($V_4$), weight SRAM ($V_5$), and convolution SRAM ($V_6$). For WG memory, only SRAMs for $s^l$ ($V_7$) and $\nabla w^l$ ($V_8$) needed new declaration, because both WG and BP share the same SRAM for $\nabla u^{l+1}$.

*B. EOCAS Overall Framework*

Fig. 3. displays the overall framework of EOCAS. The entire system takes SNN models, accelerator architecture and a memory pool as inputs to generate dataflows and evaluate the performance of each situation to obtain the optimal architecture and dataflow. For each situation, as the number and arrangement of convolution array vary, along with changes in memory types, capacities, bandwidths, bit-widths, and data types, different architectures will exhibit different performance characteristics under different dataflows.

The workload is generated based on the deep SNN models. It describes the layer, operation type, bitwidth of IOs and the loop dimensions. The layer is the number of the executing layer in selected deep SNN model. The operation type includes the following information: whether the operation is a forward pass, backward pass, or gradient update; and whether it is a floating-point or a spike-based computation. Bitwidth of IOs indicates the bitwidth of the input, the weight and the output of each operation in various layer. The loop dimension contains the orders of each convolution.

The architecture pool is generated based on the memory pool and the general accelerator architecture. In the general accelerator architecture, we define several memories to form the memory pool and the Mul-Add or Mux-Add array for convolution during training. To meet the command of certain task from workload pool, each architecture has a unique array and its corresponding memories storing inputs, weights and outputs for convolutions.

The architecture and workload are evaluated based on their derived dataflows, shown as a long loop nest with memory access information. These differences can lead to varying access requirements for a specific memory during the computation process, resulting in various performance.

*C. Performance Assessment for Convolution*

The read-write relationships among memory components are illustrated in Fig. 4. Data reuse occurs across these three storage levels, where reuse factors and unit energy for each variable at each storage level are defined as shown in Table I. The reuse factor inspired by [9] indicates the times of reading/writing from memories that can be ignored because of the maintaining property of spike-based convolution.

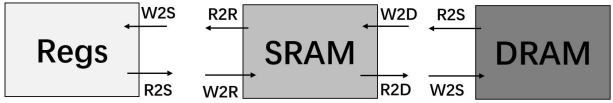

**Fig. 4.** Read/write actions at all levels of memory.

The energy consumption variables for operations of each configured memory are defined in Table II. To better describe the training process, we declare each SRAM with its bit-level energy, volume and variable bitwidth to calculate the entire energy.

Additionally, for spike convolution in FP, compared to 1-bit spikes, retaining 16-bit weights in the compute array registers clearly enhances data reuse. This type of dataflow, where weights remain stationary in the registers, is known as the Weight-Stationary (WS) dataflow. However, for floating point convolution in BP, WS may not be the most suitable one due to the accumulated sum, which possibly indicates Output-Stationary (OS) dataflow a better choice. These dataflows can be represented in the form of nested loops, where different dataflow can lead to various performance because of the amounts of reading/writing operations towards different memory.

TABLE II
ENERGY OF READ/WRITE FOR CONVOLUTIONS

| Type | Variables | Variable Bitwidth | Size | Read energy (pJ/bit) | Write energy (pJ/bit) |
|---|---|---|---|---|---|
| DRAM | - | - | - | $m_0^r$ | $m_0^w$ |
| SRAM | $s_t^{l-1}$ | 1 bit | $V_1$ | $s_0^r$ | $s_0^w$ |
| | $w^{l-1}$ | 16 bit | $V_2$ | $s_1^r$ | $s_1^w$ |
| | $ConvFP^l$ | 16 bit | $V_3$ | $s_2^r$ | $s_2^w$ |
| | $\nabla u_t^{l+1}$ | 16 bit | $V_4$ | $s_3^r$ | $s_3^w$ |
| | $w'^l$ | 16 bit | $V_5$ | $s_4^r$ | $s_4^w$ |
| | $ConvBP^l$ | 16 bit | $V_6$ | $s_5^r$ | $s_5^w$ |
| | $s_t^l$ | 1 bit | $V_7$ | $s_6^r$ | $s_6^w$ |
| | $\nabla w^l$ | 16 bit | $V_8$ | $s_7^r$ | $s_7^w$ |
| Register | *Binary* | 1 bit | - | $r_0^r$ | $r_0^w$ |
| | *Floating* | 16 bit | - | $r_1^r$ | $r_1^w$ |

Assigning the energy of Mux, Add and Multiply (Mul) operations is $o_0$, $o_1$, $o_2$ per operation, the entire computation energy of the proposed method can be written as:

$$E = E^m + E^c \quad (15)$$

where $E^c$ stands for the computing energy, and $E^m$ stands for the memory reading/writing energy.

$$E^m = E_{FP}^m + E_{BP}^m + E_{WG}^m \quad (16)$$

$$E_{FP}^C = \sum_l Mux_{ConvFP}^l \times o_0 + Add_{ConvFP}^l \times o_1 \quad (17)$$

$$E_{BP}^C = \sum_l Add_{ConvBP}^l \times o_1 + Mul_{ConvBP}^l \times o_2 \quad (18)$$

$$E_{WG}^C = \sum_l Mux_{WG}^l \times o_0 + Add_{WG}^l \times o_1 \quad (19)$$

$$E_{FP}^m = \sum_l Mux_{ConvFP}^l$$
$$\times \left( \frac{r_0^w + s_0^r}{RU_1} + \frac{s_0^w + m_0^r}{RU_2} + \frac{r_1^w + s_1^r}{RU_3} \right.$$
$$\left. + \frac{s_1^w + m_0^r}{RU_4} + \frac{r_1^r + s_2^w}{RU_5} + \frac{s_2^r + m_0^w}{RU_6} \right) \quad (20)$$

$$E_{BP}^m = \sum_l Mul_{ConvBP}^l$$
$$\times \left( \frac{r_1^w + s_3^r}{RU_7} + \frac{s_3^w + m_0^r}{RU_8} + \frac{r_1^w + s_4^r}{RU_9} \right.$$
$$\left. + \frac{s_4^w + m_0^r}{RU_{10}} + \frac{r_1^r + s_5^w}{RU_{11}} + \frac{s_5^r + m_0^w}{RU_{12}} \right) \quad (21)$$

$$E_{WG}^m = \sum_l Mux_{WG}^l$$
$$\times \left( \frac{r_1^w + s_3^r}{RU_{13}} + \frac{s_3^w + m_0^r}{RU_{14}} + \frac{r_0^w + s_6^r}{RU_{15}} \right.$$
$$\left. + \frac{s_6^w + m_0^r}{RU_{16}} + \frac{r_1^r + s_7^w}{RU_{17}} + \frac{s_7^r + m_0^w}{RU_{18}} \right) \quad (22)$$

where $Mux$, $Add$ and $Mul$ are numbers of corresponding operations mentioned previously.

*D. Analysis of Soma Unit and Grad Unit*

When both compute and memory resources are fixed, variations in dataflow have limited impact on the performance of soma and grad operations. Accordingly, their computation and memory access patterns remain stable. In our methodology, we define the memory regions associated with soma and grad units. Based on their microarchitectural design, the number of operations involved in each execution is fixed and identifiable.

At the computational level, each soma operation consists of three comparators, three multiplexers, one adder, and one multiplier. Each grad operation includes two multipliers, two adders, and two multiplexers. These components together determine the total energy cost of a single soma or grad computation.

At the memory level, the soma unit receives the forward convolution result, the membrane potential from the previous timestep, and the spike input. It produces the current timestep's spike, membrane potential, and step signal. The grad unit receives the backward convolution result, the membrane potential gradient of the next timestep, the current membrane potential, and the step signal, and outputs the current gradient of the membrane potential.

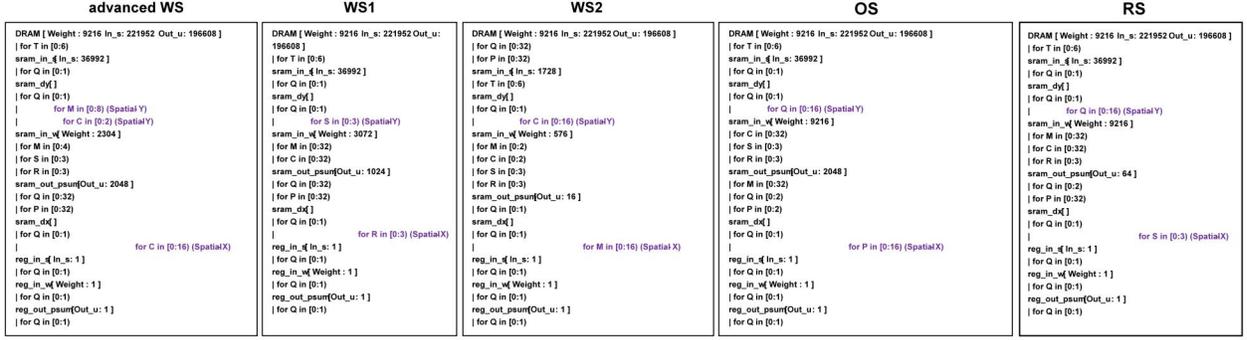

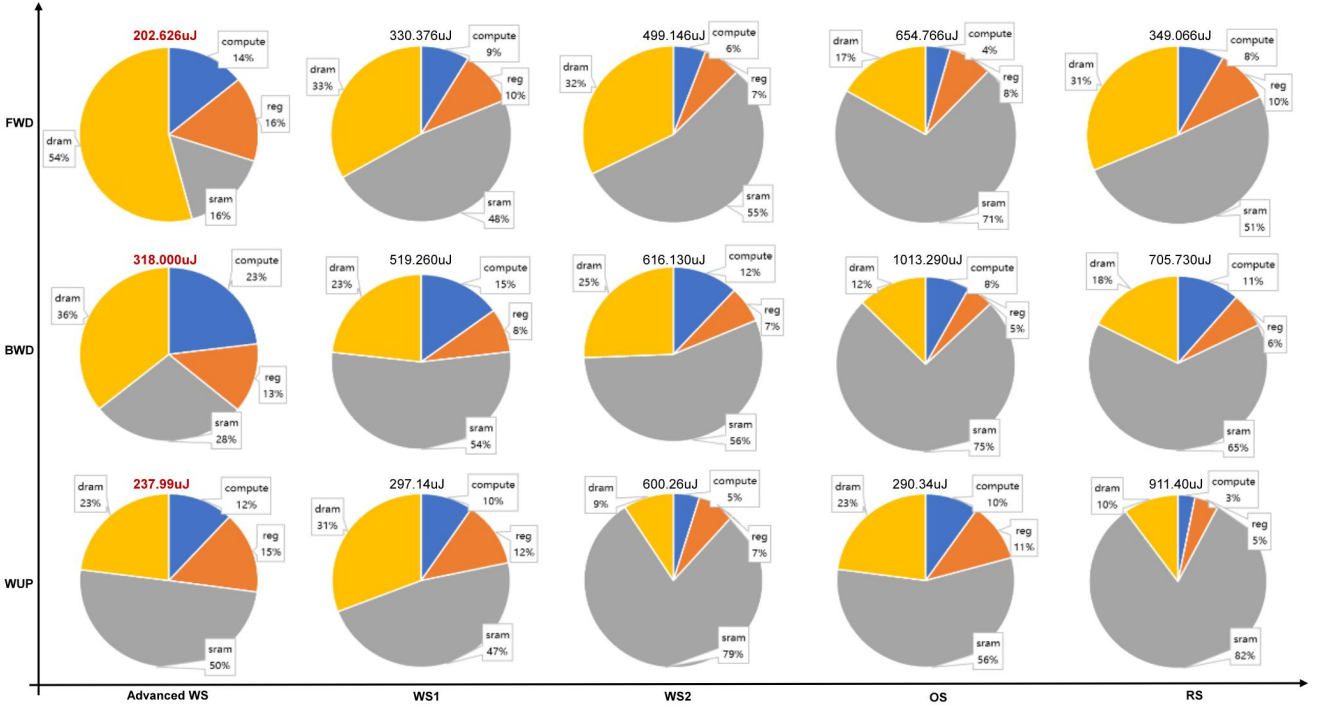

**Fig. 6.** Dataflows and the energy breakdown of convolutions using 16×16 MAC scheme.

Once the layer-wise parameters of a deep SNN are specified, the number of data transfers between corresponding SRAM modules becomes deterministic. This allows the storage energy consumption to be estimated accurately.

### IV. Experimental Results

#### A. Optimal Configurations and Dataflow

The convolutional parameters involved in the FP, BP, and WG computations of a representative layer in deep SNN training are illustrated using the CIFAR-100 dataset as an example. In the experiments, to meet high data throughput requirements, the number of MAC units is fixed at 256, leading to various array configurations such as 2×128, 4×64, 8×32, and 16×16. Several possible MAC schemes appear in different energy intervals in Fig. 5. The most reasonable array scheme is set as 16×16, yielding the optimal energy of 124.57 uJ, which is smaller than any other values during experiments in Table III.

Further, we evaluate two conventional WS dataflows, along with OS, Row-Stationary (RS) schemes and an advanced WS proposed in our experiment using the optimal configuration. The dataflow structures and their corresponding energy breakdown are detailed in Fig. 6.

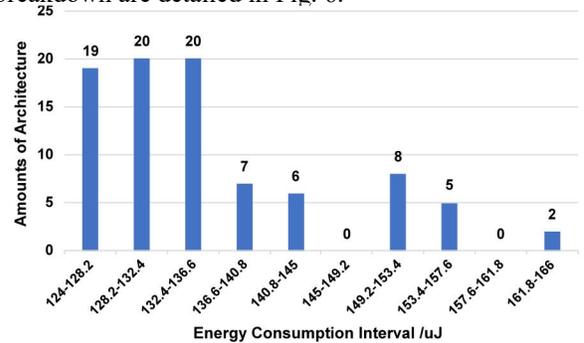

**Fig. 5.** Different energy intervals with different schemes.

In two WS dataflows, weight is preloaded and kept stationary in on-chip SRAM throughout the computation, enabling efficient weight reuse. Inputs are loaded in blocks from DRAM to SRAM in batches and processed in parallel along the vertical dimension, while partial sums along the output channel dimension are accumulated and buffered using on-chip registers and local SRAM. The entire computation is executed on-chip through a multi-level nested loop structure, starting from outer tile-level scheduling down to the inner kernel dimensions, and ultimately aligning and fusing inputs, weights, and outputs at the register level for each convolution operation.

TABLE III
ENERGY OF READ/WRITE FOR CONVOLUTIONS

| Typical Case | SRAM | MACs Amount | MACs Scheme | Energy [uJ] |
|---|---|---|---|---|
| 1 | 2.03MB | 256 | 16×16 | 124.57 |
| 2 | | | 2×128 | 156.58 |
| 3 | | | 8×32 | 141.24 |
| 4 | | | 4×64 | 135.81 |

For OS, the inputs are loaded into SRAM in blocks and then processed in parallel along the output height dimension. The weights are loaded once into SRAM and reused across multiple channels and kernel dimensions. Partial sums are buffered in partial sum SRAM and the computation is carried out across output channels and output spatial positions. The final convolution results are accumulated and buffered in registers before being written back.

For RS, similarly, all weights are loaded once into SRAM and reused across the output channels, input channels, and kernel row dimensions. Partial sums are stored row-wise in partial sum SRAM and updated along the output width dimension.

In advanced WS, input feature maps are decomposed across both output and input channel dimensions for fine-grained parallelism in the vertical direction. Weight data is partially loaded, enabling tiled reuse across output channels and spatial kernel dimensions. Partial sums are stored in partial sum SRAM and computed over the full output spatial grid, maximizing arithmetic unit utilization. Input channels are further split in the horizontal direction, ensuring scalable vector-level parallelism.

TABLE IV
OVERALL ENERGY OF DATAFLOWS

| Energy (uJ) | FP | | | BP | | | WG | | Overall |
|---|---|---|---|---|---|---|---|---|---|
| | spike conv | soma | FP total | floating point conv | grad | BP total | spike conv | WG total | |
| Advanced WS | 144.13 | 58.496 | 202.626 | 234.30 | 83.700 | 318.000 | 237.99 | 237.99 | 758.616 |
| WS1 | 271.88 | | 330.376 | 435.56 | | 519.260 | 297.14 | 297.14 | 1146.776 |
| WS2 | 440.65 | | 499.146 | 532.43 | | 616.130 | 600.26 | 600.26 | 1715.536 |
| OS | 596.27 | | 654.766 | 929.59 | | 1013.290 | 290.34 | 290.34 | 1958.396 |
| RS | 290.57 | | 349.066 | 622.03 | | 705.730 | 911.4 | 911.4 | 1966.196 |

Table IV indicates the overall energy of various dataflows, including computation and memory access. Advanced WS achieves a total energy consumption of 758.616 μJ, significantly outperforming all other methods (WS1, WS2, OS, RS) with energy reductions ranging from 33.8% to 61.4%. Specifically, it uses 33.8% less energy than WS1, 55.8% less than WS2, and approximately 61.3–61.4% less than OS and RS. The most substantial savings occur during FP and BP, driven by highly efficient spike convolution and gradient processing.

*B. Hardware Implementation Based on Optimization*

Based on the simulation and analysis of the EOCAS, the hardware architecture at the Register-Transistor-Level (RTL) implementation is in Fig. 7. Besides, we use the advanced WS dataflow to design the hardware scheme. For the convolution computation, we use the 16×16 array configuration in both FP, BP and WG. According to the general SNN training architecture, the framework is divided into two parts: the forward (FWD) and the backward (BWD). For FWD, the main operation of convolution is the Add, and the output of the Adder Matrix is stored to calculate "soma", the unit that produces the spike, compressed potential, and spike gradient mask. In BWD, both Add and MAC are conducted in BP convolution, and the outputs are finally used for "grad", the unit that produces $\nabla u_t^l$ and $\nabla s_t^l$. However, in WG, the main operation of convolution is also the Add.

This hardware architecture has been implemented on the VCU128 FPGA using Verilog HDL. The complete design was then synthesized using DC at 500 MHz with a 28nm technology library under typical conditions. Post-synthesis simulations were performed using the synthesized netlist, and performance estimations were derived through Synopsys PrimeTime. The entire power of the proposed hardware architecture is 0.452w.

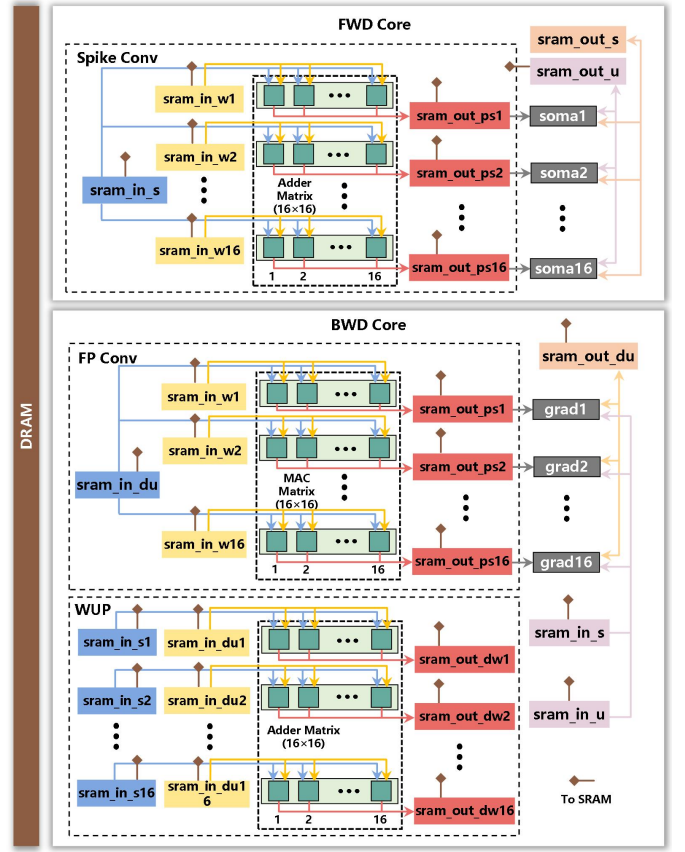

**Fig. 7.** The implemented hardware for SNN training.

To the best of our knowledge, there is currently a lack of publicly available edge devices or published hardware designs that support SNN training. Most prior works focus on inference-only accelerators. The computation of BP is more complex than FP, due to it contains the floating-point operation and the data connection in both spatial and temporal spaces. Therefore, BP utilizes more hardware resources. Compared with the FPGA-based SNN inferencing accelerators

[7] and [13] in Table IV, they can only support inferencing tasks instead of BP-based training. Thus, they need fewer LUTs, FFs, and memory usage due to their thinner computation than ours. In contrast with the DNN inferencing accelerator [14], although our design exhibits higher LUT, FF, and memory usage, it supports BP-based SNN training with reduced DSP usage.

TABLE VII
COMPARISON AMONG OTHER SOTA FPGAS

| Type | This Work | TCAS-II [7] | TCAS-II [13] | TCAS-I [14] |
|---|---|---|---|---|
| Device | VCU128 | Kintex-7 | ZCU102 | ZCU102 |
| Network | SNN | SNN | SNN | DNN |
| Training | Able | Unable | Unable | Unable |
| LUTs | 240K | 34K | 11K | 144K |
| FF | 240K | 5K | 7K | 168K |
| DSP | 1183 | 256 | - | 1268 |
| Freq (MHz) | 500 | 143 | 200 | 300 |

Compared to the ASIC (Application Specific Integrated Circuit) SNN inferencing work proposed by F. Akopyan [4] with multiple cores, our design demonstrates advantages in memory usage and BP computation. Moreover, even if their design has smaller power consumption, our energy efficiency is 2.76× higher than theirs. Besides, our work achieves a 49.25% lower memory usage against SNN training-supported design SATA [15]. Compared to efficient training architecture [16] for Transformer [17], our weight precision is 2× wider than their 8-bit PINT (8,3) format. Their efficiency is higher than ours, due to the strong performance of Transformer attention mechanisms. However, our energy consumption is approximately one-tenth of theirs, demonstrating the low power of SNN computation.

TABLE VII
COMPARISON AMONG OTHER SOTA ASICS

| Type | This Work | TCAD [4] | TCAD [15] | TVLSI[16] |
|---|---|---|---|---|
| Process | 28nm | 28nm | 65nm | 28nm |
| Network | SNN | SNN | SNN | DNN |
| Training | Able | Unable | Unable | Able |
| Weight Precision | FP16 | INT1 | INT8 | PINT(8,3) |
| Memory (MB) | 2.03 | - | 4 | - |
| Throughputs(TOPS) | 0.5 | 0.0581 | - | 14.71 |
| Area (mm$^2$) | 6.83 | 430 | - | 17.26 |
| Power (w) | 0.452 | 0.065 | - | 4.45 |
| Energy Efficiency (TOPS/W) | 1.11 | 0.4 | - | 3.31 |

V. CONCLUSION

This paper presents an Energy-Oriented Computing Architecture Simulator (EOCAS) for SNN training. EOCAS defines the hardware design space representation, assesses energy, and selects the optimal architecture. Under the guidance of EOCAS, we implement the power-aimed optimal hardware architecture and achieve low energy consumption. Compared with several SOTA DNN and SNN works, our hardware architecture outstands others in various criteria.

REFERENCES

[1] Deepak, M. K. Upadhyay and M. Alam, "Edge Computing: Architecture, Application, Opportunities, and Challenges," *2023 3rd International Conference on Technological Advancements in Computational Sciences (ICTACS)*, Tashkent, Uzbekistan, 2023, pp. 695-702.
[2] M. Dampfhoffer, T. Mesquida, A. Valentian and L. Anghel, "Backpropagation-Based Learning Techniques for Deep Spiking Neural Networks: A Survey," in *IEEE Transactions on Neural Networks and Learning Systems*, vol. 35, no. 9, pp. 11906-11921, Sept. 2024.
[3] Wolfgang Maass, Networks of spiking neurons: The third generation of neural network models, *Neural Networks*, Volume 10, Issue 9, 1997, Pages 1659-1671, ISSN 0893-6080.
[4] F. Akopyan et al., "TrueNorth: Design and Tool Flow of a 65 mW 1 Million Neuron Programmable Neurosynaptic Chip," in *IEEE Transactions on Computer-Aided Design of Integrated Circuits and Systems*, vol. 34, no. 10, pp. 1537-1557, Oct. 2015.
[5] M. Davies et al., "Advancing Neuromorphic Computing With Loihi: A Survey of Results and Outlook," in *Proceedings of the IEEE*, vol. 109, no. 5, pp. 911-934, May 2021.
[6] Pei, J., Deng, L., et al. Towards artificial general intelligence with hybrid tianjic chip architecture. *Nature*, 572, 106–111(2019).
[7] H. Asgari, B. M. -N. Maybodi, M. Payvand and M. R. Azghadi, "Low-Energy and Fast Spiking Neural Network For Context-Dependent Learning on FPGA," in *IEEE Transactions on Circuits and Systems II: Express Briefs*, vol. 67, no. 11, pp. 2697-2701, Nov. 2020.
[8] A. Parashar et al., "Timeloop: A Systematic Approach to DNN Accelerator Evaluation," *2019 IEEE International Symposium on Performance Analysis of Systems and Software (ISPASS)*, Madison, WI, USA, 2019, pp. 304-315.
[9] L. Mei, P. Houshmand, V. Jain, S. Giraldo and M. Verhelst, "ZigZag: Enlarging Joint Architecture-Mapping Design Space Exploration for DNN Accelerators," in *IEEE Transactions on Computers*, vol. 70, no. 8, pp. 1160-1174, 1 Aug. 2021.
[10] L. Liang et al., "H2Learn: High-Efficiency Learning Accelerator for High-Accuracy Spiking Neural Networks," in *IEEE Transactions on Computer-Aided Design of Integrated Circuits and Systems*, vol. 41, no. 11, pp. 4782-4796, Nov. 2022.
[11] Y. Li, M. Wen, R. Yang, J. Shen, Y. Cao and J. Wang, "S-SIM: A Simulator for Systolic Array-based DNN Accelerators with Tile Access Awareness," *2022 IEEE International Symposium on Circuits and Systems (ISCAS)*, Austin, TX, USA, 2022, pp. 2720-2724.
[12] X. Yi et al., "NNASIM: An Efficient Event-Driven Simulator for DNN Accelerators with Accurate Timing and Area Models," *2022 IEEE International Symposium on Circuits and Systems (ISCAS)*, Austin, TX, USA, 2022, pp. 2806-2810.
[13] H. Zheng, Y. Guo, X. Yang, S. Xiao and Z. Yu, "Balancing the Cost and Performance Trade-Offs in SNN Processors," in *IEEE Transactions on Circuits and Systems II: Express Briefs*, vol. 68, no. 9, pp. 3172-3176, Sept. 2021.
[14] M. Huang, J. Luo, C. Ding, Z. Wei, S. Huang and H. Yu, "An Integer-Only and Group-Vector Systolic Accelerator for Efficiently Mapping Vision Transformer on Edge," in *IEEE Transactions on Circuits and Systems I: Regular Papers*, vol. 70, no. 12, pp. 5289-5301, Dec. 2023.
[15] R. Yin, A. Moitra, A. Bhattacharjee, Y. Kim and P. Panda, "SATA: Sparsity-Aware Training Accelerator for Spiking Neural Networks," in *IEEE Transactions on Computer-Aided Design of Integrated Circuits and Systems*, vol. 42, no. 6, pp. 1926-1938, June 2023.
[16] H. Shao, J. Lu, M. Wang and Z. Wang, "An Efficient Training Accelerator for Transformers With Hardware-Algorithm Co-Optimization," in *IEEE Transactions on Very Large Scale Integration (VLSI) Systems*, vol. 31, no. 11, pp. 1788-1801, Nov. 2023, doi: 10.1109/TVLSI.2023.3305569.
[17] Vaswani A, Shazeer N, Parmar N, et al. Attention is all you need[J]. Advances in neural information processing systems, 2017, 30.
[18] H. Zhang, A. Ning, R. B. Prabhakar and D. Wentzlaff, "LLMCompass: Enabling Efficient Hardware Design for Large Language Model Inference," *2024 ACM/IEEE 51st Annual International Symposium on Computer Architecture (ISCA)*, Buenos Aires, Argentina, 2024, pp. 1080-1096, doi: 10.1109/ISCA59077.2024.00082.
[19] S. Yang et al., "Scalable Digital Neuromorphic Architecture for Large-Scale Biophysically Meaningful Neural Network With Multi-Compartment Neurons," in *IEEE Transactions on Neural Networks and Learning Systems*, vol. 31, no. 1, pp. 148-162, Jan. 2020, doi: 10.1109/TNNLS.2019.2899936.

[20] S. Yang *et al*., "BiCoSS: Toward Large-Scale Cognition Brain With Multigranular Neuromorphic Architecture," in *IEEE Transactions on Neural Networks and Learning Systems*, vol. 33, no. 7, pp. 2801-2815, July 2022, doi: 10.1109/TNNLS.2020.3045492.

[21] Y. -H. Chen, T. Krishna, J. S. Emer and V. Sze, "Eyeriss: An Energy-Efficient Reconfigurable Accelerator for Deep Convolutional Neural Networks," in *IEEE Journal of Solid-State Circuits*, vol. 52, no. 1, pp. 127-138, Jan. 2017, doi: 10.1109/JSSC.2016.2616357.